\documentstyle[aps,epsf,preprint]{revtex}

\newcommand{\ts}{\textstyle}
\newcommand{\ds}{\displaystyle}

\tightenlines
\preprint{RCNP--Th01015}
\title{\bf The Influence of an External Chromomagnetic Field on
Color Superconductivity}
\author{\bf D.~Ebert}
\address{ Research
Center for Nuclear Physics (RCNP), Osaka University, Ibaraki,Osaka
567,Japan}
\address{and Institut f\"ur Physik,
 Humboldt-Universit\"at zu Berlin, 10115 Berlin, Germany}

\author{\bf V.V.~Khudyakov and V.Ch.~Zhukovsky}
\address{Faculty of
Physics, Department of Theoretical Physics, Moscow State University,
119899, Moscow, Russia}

\author{\bf K.G. Klimenko}
\address{Institute
of High Energy Physics, 142284, Protvino, Moscow Region, Russia}

\begin{document}
\draft
\large
\maketitle

\begin{abstract}
We study the competition of quark-antiquark and
diquark condensates under the influence of an external chromomagnetic
field modelling the gluon condensate
and in dependence on the chemical potential and temperature.
As our results indicate,   
an external chromomagnetic field
might produce remarkable
qualitative changes in the picture of the
color superconducting (CSC) phase formation. 
This concerns, in particular, the possibility of a transition to
the CSC phase and diquark condensation at finite temperature.

\end{abstract}

\newcommand\tg{\mathop{\rm tg}}
\newcommand\const{\mathop{\rm const}}
\newcommand\mvec[1]{{\mathbf{.)1}}}
\newcommand\op[1]{\mathop {\rm#1}\nolimits}
\newcommand\up[1]{{\rm#1}}
\newcommand\ch{\mathop{\rm ch}}
\newcommand\th{\mathop{\rm th}}
\newcommand\wt{\widetilde}
\arraycolsep=1.8pt
\font\eightrm=cmr8

\vspace{0.3cm}
\large

\renewcommand{\thefootnote}{\arabic{footnote}}
\setcounter{footnote}{0}
\setcounter{page}{1}

\section{Introduction}

Low energy (large distance)  effects in 
QCD can only be studied by approximate (nonperturbative)
methods in the framework of various effective models
or in terms of lattice calculations.
At present time, one of the most popular QCD-like 
effective theories is the well-known Nambu--Jona-Lasinio (NJL) model
\cite{njl}, which is 
a relativistic quantum field theory
with four-fermion interactions.
The physics of
light mesons (see e.g. \cite {7} and references therein),
diquarks \cite {ebertt,vog} and meson-baryon interactions 
\cite {ebjur}-\cite {ishi}
based on dynamical chiral symmetry breaking
can be effectively described by NJL chiral
quark models. Moreover, NJL models
are widely  used in nuclear
physics and astrophysics (neutron stars) for the 
investigation of quark
matter \cite{satar}, 
to construct alternative models of electroweak interactions
\cite{w} and in cosmological applications \cite{odin}.
Moreover, its (2+1)-dimensional analogue serves as a satisfactory
microscopic theory for several effects in the physics 
of high-temperature superconductors \cite{liu}.
 
The NJL model displays the same symmetries as QCD. So it can be 
successfully used for simulating some of the QCD vacuum
properties under the influence of external conditions such as
temperature $T$ and chemical potential $\mu$ \cite{kawati}. The
role
of such considerations significantly increases 
especially in the cases, where numerical lattice calculations are 
not admissible in QCD, i.e. at nonzero density and in the 
presence of external electromagnetic fields \cite{kl,ekvv}. 
Recently, it was
shown in the framework of a (2+1)-dimensional NJL model that an
arbitrary small external magnetic field induces the spontaneous
chiral symmetry breaking ($\chi$SB)
even under conditions, when the interaction between fermions is
arbitrary weak \cite{2}. Later it was shown that this phenomenon
(called magnetic catalysis effect) has a rather universal character
and
gets its explanation on the basis of the dimensional
reduction mechanism \cite{gus}. (The recent reviews \cite{gus2}
consider the modern status of
the magnetic catalysis effect and its applications in
different branches of physics.)

As an effective theory for low energy QCD, the NJL model does
not contain any dynamical gluon fields. Such nonperturbative
feature of the real QCD vacuum, as the nonzero gluon condensate
$<F^a_{\mu\nu}F^{a\mu\nu}>\equiv <FF>$ can, however, 
be mimicked in the framework of 
NJL models with the help of external chromomagnetic fields.
In particular, for a 
QCD-motivated NJL model with gluon condensate (i.e. in the 
presence of an external chromomagnetic field) and finite
temperature, it was shown that a weak gluon condensate plays a
stabilizing role for the behavior of the constituent quark mass,
the quark condensate, meson masses and coupling constants for
varying temperature \cite{8}. Then, in a series of papers,
devoted to the NJL model with gluon condensate, it was 
shown that an external chromomagnetic field, similar to the ordinary
magnetic field, 
serves as a catalyzing factor in the fermion mass generation
and dynamical breaking of chiral symmetry as well \cite{klim}.
The basis for this phenomenon is the effective reduction of the 
space dimensionality in the presence of external chromomagnetic
fields \cite {zheb}.

There exists the exciting idea proposed more than twenty years 
ago \cite{Ba}-\cite{bl} that at high baryon densities a colored
diquark condensate $<qq>$ might appear.
 In analogy with ordinary
superconductivity, this effect was called color superconductivity
(CSC). 
The CSC phenomenon was investigated in the framework of the one-gluon
exchange approximation in QCD \cite{son}, where
the colored Cooper pair formation is predicted selfconsistently
at extremely high values of the chemical potential $\mu\gtrsim 10^8$
MeV \cite{raj}. Unfortunately, such baryon densities are not
observable in nature and not accessible in experiments (the typical
densities inside the neutron stars or in the future heavy ion
experiments correspond to  $\mu\sim 500$ MeV). The
possibility for the existency of the CSC phase in the region
of moderate densities was proved quite recently
(see e.g. the papers \cite{rapp}-\cite{klev}
as well as the review articles \cite{alf2} and references therein).
In these papers it was shown on the basis of different effective
theories for low energy QCD (instanton model, NJL model, etc.)
that the 
diquark condensate $<qq>$ can appear  already at a rather moderate
baryon density ($\mu\sim 400$ MeV), which can possibly be
detected in the future experiments on ion-ion collisions.
Since quark Cooper pairing occurs in the color anti-triplet
channel, the
nonzero value of $<qq>$ means that, apart from
the electromagnetic $U(1)$ symmetry,  the color
$SU_c(3)$ should be spontaneously broken down  inside
the CSC phase as well. 
In the framework of NJL models the CSC phase formation has generally 
be considered
as a dynamical competition between diquark  $<qq>$ and
usual quark-antiquark condensation $<\bar qq>$.
However, the real QCD vacuum is characterized in addition by the
appearence
of a gluon condensate $<FF>$ as well, which might 
change the generally 
accepted conditions for the CSC observation. In particular,
one would expect that, similarly to the case of  quark-antiquark
condensation, the process of
diquark condensation might be induced by external
chromomagnetic fields. For a (2+1)-dimensional quark model, this was
recently demonstrated in \cite {ebklim}.
There, a $SU(2)_L\times SU(2)_R$ 
chirally symmetric (2+1)-dimensional NJL model with three colored
quarks of two flavors was considered at zero $T,\mu$. It was shown 
that in this case for arbitrary fixed values of coupling constants
there exists a critical value of the external chromomagnetic field
at which a CSC second order phase transition is induced in the
system.
\footnote{Strictly speaking, the CSC is induced by those components
of external chromomagnetic fields, which can stay massless inside the
CSC phase.}
Since the two-flavored QCD$_3$ and the considered NJL model are not
in the same 
universality class of theories (QCD$_3$ with N$_f=2$ has a higher
flavor symmetry $SU(4)$), the obtained results are intrinsic to 
real QCD$_4$ rather than to QCD$_3$.
Indeed, our recent investigations on the basis of a (3+1)-dimensional
NJL model \cite{gap,gap1} and $\mu=0$ show that some types of
sufficiently
strong external chromomagnetic fields may catalyze the diquark
condensation.

As argued above, CSC might occur
inside neutron stars and possibly become observable in ion-ion
collisions,
i.e. at nonzero baryon densities. 
Taking into account the fact that at finite chemical potential
the magnetic generation of dynamical $\chi$SB 
qualitatively differs from the $\mu=0$ case \cite{ekvv},
one might expect analogous effects for CSC, too. By this reason,
the investigation of the chromomagnetic generation of CSC under the
influence of a finite chemical potential (finite particle density) 
is a very interesting and actual physical problem.

The aim of the present paper is to study the influence of external
conditions such as 
chemical potential, temperature  and especially  of the gluon
condensate (as modelled by external color gauge fields) on the phase
structure of quark matter with particular emphasize of its CSC phase.
To this end, we shall extend our earlier
analysis of the 
chromomagnetic generation of CSC at $\mu=0$ \cite
{ebklim}-\cite{gap1}
to the case of an
(3+1)-dimensional  NJL type model  with finite chromomagnetic field, 
temperature and chemical potential presenting a generalization of the
free field model of \cite{klev}.
  
The paper is organized as follows. In Sections II and III the
extended
NJL model under consideration is presented, and its effective
potential ($\equiv$ thermodynamic potential) at nonzero external
chromomagnetic field, chemical potential and temperature is obtained
in the one-loop approximation. This quantity contains all the
necessary informations about the quark and diquark condensates of the
theory. In the following Section IV  
the phase structure of the model is discussed on the basis of 
numerical investigations of the global minimum point of the 
effective potential.
As our main result, it is shown that the external chromomagnetic
field can induce the transition
to the CSC phase and diquark condensation
even at finite temperature.
Thereby, 
the characteristics of the CSC phase can significantly 
change in dependence on the strength of the chromomagnetic field. 
Finally, Section V contains 
a summary and discussion of the results. 
Some details of the effective potential calculation are relegated to
an Appendix.

\section{The model}

Let us first give several (very approximative) arguments motivating
the chosen
structure of our QCD-motivated extended NJL model introduced below.
For this aim, consider two-flavor QCD with nonzero chemical
potential and 
color group 
$SU_c(N_c)$ and decompose the gluon field ${\cal
A}_\nu^a(x)$
into a condensate background (``external'') field $A_\nu^a(x)$ and
the 
quantum fluctuation $a_\nu^a(x)$ around it, i.e. 
${\cal A}_\nu^a(x)=$$A_\nu^a(x)+$$a_\nu^a(x).$
By integrating in the generating functional of QCD over
the quantum field $a_\nu^a(x)$ and further ``approximating'' the 
nonperturbative gluon propagator by a $\delta-$function,
one arrives at an effective local chiral four-quark interaction of
the
NJL type describing low energy hadron physics 
in the presence of a gluon condensate. Finally, by performing a
Fierz transformation of the interaction term, one obtains 
a four-fermionic model with $(\bar q q)$--and $(q q)$--interactions
and an external condensate field $A_\mu^a(x)$
of the color group $SU_c(N_c)$  given by the following
Lagrangian
\footnote{The most general four-fermion interaction would
include additional vector and axial-vector $(\bar q q)$ as
well as pseudo-scalar, vector and axial-vector-like $(q q)$
-interactions. 
For our goal of studying the effect of
chromomagnetic
catalysis for the competition of quark and diquark condensates, the
interaction structure of (1) is, however, sufficiently general.}
\begin{eqnarray}
  L&=&\bar q[\gamma^\nu(i\partial_\nu+g A_\nu^a(x)\frac{\lambda^a}2)
+  \mu\gamma^0]q+\frac{G_1}{2N_c}[(\bar qq)^2+(\bar qi\gamma^5\vec
  \tau q)^2]+\nonumber\\
  &+&\frac{G_2}{N_c}[i\bar q_c\varepsilon
(i\lambda_{as}^b)       \gamma^5 q]
  [i\bar q\varepsilon
(i\lambda_{as}^b)\gamma^5 q_c].
  \label{x1}
\end{eqnarray}
It is necessary to note that in order to obtain 
realistic estimates for masses of 
vector/axial-vector mesons and diquarks 
in extended NJL--type of models \cite{ebertt}, we have to allow for
independent coupling constants 
$G_1, G_2$, rather than to consider them related by a Fierz
transformation
of a current-current interaction via gluon exchange. Clearly, such a  
procedure does not spoil chiral symmetry. 

In (\ref{x1}) $g$ denotes the gluon coupling constant,
$\mu$ is the quark chemical potential, $q_c=C\bar
q^t$, $\bar q_c=q^t C$ are charge-conjugated spinors,
and $C=i\gamma^2\gamma^0$ 
is the charge conjugation matrix ($t$ denotes
the transposition operation). 
In what follows we assume $N_c=3$ 
and replace the antisymmetric color matrices
$\lambda_{as}^b$ (with a factor $i$) by the antisymmetric
$\epsilon^b$
operator. 
Moreover, summation over repeated color indices $a =1,\dots,8$; $b =
1,2,3$ and Lorentz indices
$\nu = 0,1,2,3$ is implied. The quark field $q\equiv q_{i\alpha}$ is
a flavor doublet and color triplet as well as a four-component Dirac
spinor, where $i=1,2$; $\alpha = 1,2,3$. (Latin and Greek indices
refer to flavor and color 
indices, respectively; spinor indices are
omitted.) Furthermore, we use the notations $\lambda^a/2$ for the
generators of the
color $SU_c(3)$ group 
appearing in the covariant derivative 
as well as $\vec \tau\equiv (\tau^{1},
\tau^{2},\tau^3)$ for Pauli
matrices in the flavor space; 
$(\varepsilon)^{ik}\equiv\varepsilon^{ik}$,
$(\epsilon^b)^{\alpha\beta}\equiv\epsilon^{\alpha\beta b}$
are totally antisymmetric tensors in the flavor and color spaces,
respectively. Clearly, the Lagrangian (\ref{x1}) is invariant under
the chiral $SU(2)_L\times SU(2)_R$ and color $SU_c(3)$ groups.

Next, let us for a  moment suppose that in (\ref{x1}) 
$A_\mu^a(x)$ is an arbitrary
classical gauge field of the color group $SU_c(3)$. (The following
investigations do not require the explicit inclusion of the gauge
field part
of the Lagrangian). The detailed structure of $A_\mu^a(x)$
corresponding
to a constant chromomagnetic gluon condensate will be given below.

The linearized version of the model (\ref{x1}) with auxiliary bosonic
fields has
the following form
\begin{eqnarray}
\tilde L\ds &=&\bar q[\gamma^\nu(i\partial_\nu+g
 A_\nu^a(x)\frac{\lambda_a}2)+\mu\gamma^0]q
 -\bar q(\sigma+i\gamma^5\vec
 \tau\vec\pi)q-\frac{3}{2G_1}(\sigma^2+\vec \pi^2)-
 \nonumber\\
&-&\frac3{G_2}\Delta^{*b}\Delta^b-
\Delta^{*b}[iq^tC\varepsilon\epsilon^b\gamma^5 q]
 -\Delta^b[i\bar q \varepsilon\epsilon^b\gamma^5 C\bar q^t].
 \label{x2}
\end{eqnarray}

The Lagrangians (\ref{x1}) and (\ref{x2}) are equivalent, as can be
seen by using the
equations of motion for bosonic fields, from which it follows that
\begin{equation}
  \Delta^b\sim iq^t C\varepsilon \epsilon^b\gamma^5 q,\quad
  \sigma\sim\bar qq,\quad
  \vec \pi\sim i\bar q\gamma^5\vec\tau q.
\end{equation}

Clearly, the $\sigma$ and $\vec\pi$ fields are color singlets.
Besides, the (bosonic)
diquark field $\Delta^{b}$ is a color antitriplet and a (isoscalar)
singlet
under the chiral
$SU(2)_L\times SU(2)_R$ group. Note further that the $\sigma$,
$\Delta^b$, are scalars, but
the $\vec\pi$ are pseudo-scalar fields. Hence, if $\sigma\ne0$, then
chiral
symmetry of the model is spontaneously broken, whereas $\Delta^b\ne0$
indicates the dynamical
breaking of both the color and electromagnetic  symmetries of the
theory.

In the one-loop approximation, the 
effective action for the boson fields which is invariant 
under the chiral (flavor) as well as color and Lorentz groups
is expressed
through the path integral over quark fields:
$$
  \exp(i S_{\up{eff}}(\sigma,\vec\pi,\Delta^b,\Delta^{*b},A_\mu^a))=
  N'\int[d\bar q][dq]\exp\Bigl(i\int\tilde L\,d^4 x\Bigr),
$$
where
\begin{equation}
  S_{\up{eff}}(\sigma,\vec\pi,\Delta^b,\Delta^{*b},A_\mu^a)=-N_c\int
  d^4x
  \left[\frac{\sigma^2+\vec\pi^2}{2G_1}+\frac{\Delta^b\Delta^{*b}}{G_
  2}\right]+\tilde S,
  \label{x4}
\end{equation}
$N'$ is a normalization constant.
The quark contribution to the partition function is 
here given by:
\begin{equation}
  Z_q=\exp(i\tilde S)=N'\int[d\bar q][dq]\exp\Bigl(i\int[\bar q{\cal
  D}q+
  \bar q{\cal M}\bar q^t+q^t\bar{\cal M}q]d^4 x\Bigr).
\end{equation}
In (5) we have used the following notations
$$
  {\cal D}=D+\gamma^\mu
  gA_\mu^a(x)\frac{\lambda^a}2;\qquad
  D=i\gamma^\mu\partial_\mu-\sigma-i\gamma^5\vec\pi\vec\tau+\mu\gamma
  ^0,
$$
\begin{equation}
  \bar{\cal M}=-i\Delta^{*b}C\varepsilon\epsilon^b\gamma^5,\qquad
  {\cal M}=-i\Delta^{b}\varepsilon\epsilon^b\gamma^5C,\qquad
\end{equation}
where $D$ is the Dirac operator in the coordinate, spinor and flavor
spaces, whereas
${\cal D}$, ${\cal M}$ and $\bar{\cal M}$ are in addition
operators
in the color space, too.
Let us next assume that in the ground state of our model
$\langle\Delta^1\rangle=
\langle\Delta^2\rangle=\langle\vec\pi\rangle=0$ and
$\langle\sigma\rangle$, ${\langle\Delta^3\rangle\ne0}$.
\footnote{If
$\langle\vec\pi\rangle\ne0$ then one would have spontaneous breaking
of parity.
For strong interactions parity is, however, a conserved
quantum number, justifying the assumption
$\langle\vec\pi\rangle=0$.}
Obviously, the residual symmetry group of such a vacuum
is $SU_c(2)$ whose generators are the first three generators of
the initial $SU_c(3)$.
Now suppose that in this frame the constant external chromomagnetic
field,
simulating the presence of a gluon condensate
$\langle FF\rangle=2H^2$,
has the following form $H^a=(H^1, H^2, H^3,0,\dots,0)$. Furthermore,
due to the residual $SU_c(2)$ invariance of the vacuum, 
one can put $H^{1}= H^2=0$ and
$H^{3}\equiv H.$

Some remarks about the structure of the external chromomagnetic  
fields
$A_\nu^a(x)$ used in (\ref{x1}) are needed. From this moment on, we
assume $A_\nu^a(x)$ in such a form that the only nonvanishing
components of
the
corresponding field strength tensor $F_{\mu\nu}^a$ are
$F_{12}^3=-F_{21}^3=H=\up{const}$.
The above homogeneous chromomagnetic field can be generated by the
following vector-potential
\begin{equation}
  A_\nu^3(x)=(0,0,Hx^1,0);\quad A_\nu^a(x)=0 \quad (a\ne 3),
  \label{x7}
\end{equation}
which defines the well known Matinyan--Savvidy model of the gluon
condensate in $QCD$ \cite{savvidy}.

In $QCD$ the physical vacuum may be interpreted as a region splitted
into an
infinite number of domains with macroscopic extension \cite{nielsen}.
Inside each such domain there can be excited a homogeneous background 
chromomagnetic field, which generates a nonzero gluon condensate
$\langle FF\rangle\ne0$.  (Averaging over all domains results in a
zero
background chromomagnetic field, hence color as well as Lorentz
symmetries are not broken.) \footnote{Strictly speaking, our
following
calculations refer to some given macroscopic domain.  The obtained
results turn out to depend on color and rotational (Lorentz)
invariant quantities
only, and are independent on the concrete domain.}

In order to find nonvanishing condensates $\langle\sigma\rangle$ and
$\langle\Delta^3\rangle$, we should calculate the effective
potential,
whose global minimum point provides us with these quantities.
Suppose that (apart from the external vector-potential $A_\mu^a(x)$
(\ref{x7}))
all boson fields in $S_{\up{eff}}$ (\ref{x4}) do not depend on
space-time.
In this case, by definition, $S_{\up{eff}}=-V_{\up{eff}}\int d^4x$,
where
\begin{eqnarray}
\label{eq.8}
V_{\rm eff}=\frac{3(\sigma^2+\vec\pi^2)}{2G_1}+\frac
{3\Delta^b\Delta^{\ast b}}{G_2}+\tilde V;~
~\tilde V=-\frac{\tilde S}{v},~~v=\int d^4x.
\end{eqnarray}

Due to our assumption on the vacuum structure,
we put $\Delta^{1,2}\equiv 0$, as well as $\vec\pi = 0$. Then, taking
into
account the form of the vector-potential (7), one can
easily see that the functional integral for $\tilde S$ in (5) is
factorized
\begin{eqnarray}
  \label{kx9}
 Z_q=\exp(i\tilde S(\sigma,\Delta))=N'\int[d\bar
  q_3][dq_3]\exp\Bigl(i\int
  \bar q_3\tilde D q_3d^4x\Bigr)\times\\
  \times \int[d \bar Q][dQ]\exp\Bigl(i\int[\bar Q\tilde{\cal D}Q+\bar
  Q M\bar Q^t+
  Q^t\bar MQ]d^4x\Bigr),
  \label{x9}
\end{eqnarray}
where $\Delta\equiv\Delta^3$, $q_3$ is the quark field of color 3 and
$Q\equiv(q_1,q_2)^t$ is the doublet, composed from quark fields of
the
colors 1,2. Moreover,
$\tilde D=D|_{\vec\pi=0}$ ($D$ is presented in (6)) and
\begin{equation}
  \tilde{\cal D}=\tilde D+\gamma^\mu
  gA_\mu^3(x)\frac{\sigma_3}2;\quad
  \bar M=-i\Delta^* C\varepsilon\tilde\epsilon\gamma^5,\quad
  M=-i\Delta\varepsilon\tilde\epsilon\gamma^5 C.
  \label{x10}
\end{equation}
In (\ref{x10}) 
$\sigma_3$, $\tilde \epsilon$ are matrices in the two-dimensional
color subspace, corresponding to the $SU_c(2)$ group:
$$
\sigma_3=
\left (\begin{array}{cc}
1 & 0\\
0 &-1
\end{array}\right ),\qquad
\tilde\epsilon=
\left (\begin{array}{cc}
0 & 1\\
-1 &0
\end{array}\right ).
$$
Clearly, the integration over $q_3$ in (\ref{kx9}) yields
$\op{det}\tilde D$.

Defining $\Psi^t=(Q^t, \bar{Q})$ and introducing the matrix-valued
operator 
$$
Z=
\left (\begin{array}{cc}
2\bar M~, & -\tilde{\cal D}^t\\
\tilde{\cal D}~, &2M
\end{array}\right ),
$$
the gaussian integral over $\bar Q$ and $Q$ in (\ref{x9}) can be
rewritten in compact matrix notation and be evaluated as 
\begin{equation}
  \int[d\Psi]  e^{\frac i2\int\Psi^t Z\Psi d^4x}=\op{det}^{1/2}Z.
\label{90}
\end{equation}
Then, by using in (\ref{90}) the general formula
$$
\det\left
(\begin{array}{cc}
A~, & B\\
\bar B~, & \bar A
\end{array}\right )=\det [-\bar BB+\bar BA\bar B^{-1}\bar A]=\det
[\bar AA-\bar AB\bar A^{-1}\bar B],
$$
one obtains the result:
\begin{eqnarray}
\exp (i\tilde S(\sigma,\Delta)) &=&
N'{\rm det}(\tilde D){\rm det}^{1/2}[4M\bar M+ M\tilde {\cal D}^t
M^{-1}\tilde{\cal D}]=\nonumber\\&=&N'{\rm
det}[(i\hat\partial-\sigma+\mu\gamma^0)]\cdot\nonumber\\
\cdot{\rm det}^{1/2}\left [4|\Delta|^2
\right.&+&(\left.-i\hat\partial-\sigma+\mu\gamma^0-g\hat A^3
\frac{\sigma_3}2)
(i\hat\partial-\sigma+\mu\gamma^0+g\hat A^3\frac
{\sigma_3}2)\right ].
\label{kp3_12}
\end{eqnarray}
Recall that the operator under the first ${\rm det}$-symbol
in (\ref{kp3_12}) acts only in the flavor,
coordinate and spinor
spaces, whereas the operator under the second
${\rm det}$-symbol acts in the two-dimensional color subspace,
too.

\section{The general case $\mu \ne 0, T \ne 0, H\ne 0$}
\subsection{The effective potential}

First of all, let us calculate the effective action from
(\ref{kp3_12}) at zero temperature $T$.
It is convenient to rewrite the second
determinant in (\ref{kp3_12}) in the form
\begin{eqnarray}
&&{\rm det}[4M\bar M+M\tilde {\cal D}^t(
M)^{-1}\tilde{\cal D}]=\nonumber\\
&=&{\rm det}\left [4|\Delta|^2
+\mu^2-p^2_0+\sigma^2-(\bar\gamma\bar\nabla)^2-
2\mu\gamma^0(\sigma+\bar\gamma\bar\nabla)\right ]
\label{eq.14}
\end{eqnarray}
where the
$p^0$-momentum space representation and $\bar\gamma\bar\nabla$ =
$\gamma_k(i\partial_k+g A^3_k\sigma_3/2)$, ($k=1,2,3$) have been
used.
Similarly to quantum electrodynamics, it is easily seen that
the operator ${\cal H}\equiv\gamma^0(\sigma+\bar\gamma\bar\nabla)$ is
the
Hamiltonian for quarks with color indices $\alpha=1,2$ and flavor
$i=1,2$ in
the background
vector-potential (\ref{x7}). Its eigenvalues are
$\pm\varepsilon_{\{n\}}$, where
$\varepsilon_{\{n\}}=\sqrt{\sigma^2+p_3^2+gH(n+1/2)-
gH\zeta/2}$, and corresponding eigenstates are denoted by
$\Phi^{\pm}_{\{n\}p_2i\alpha}$.
The set of quark quantum numbers in the background field
are defined as follows:
$\{n\}\equiv\{n=0,1,2,...;~-\infty<p_3<+\infty;~\zeta=\pm 1\}$,
$i,\alpha =1,2$,~$-\infty<p_2<\infty$.
Each of the
eigenvalues $\pm\varepsilon_{\{n\}}$ for ${\cal H}$ is evidently
four-fold
degenerate with respect to flavor and color quantum numbers
$i,\alpha$. It is also
degenerate with respect to the quantum number $p_2$, which
quasiclassically characterizes the charged
particle center of orbit
position in an external uniform  magnetic field.  Since \[ {\cal
H}{\cal
H}\Phi^{\pm}_{\{n\}p_2i\alpha}=[\sigma^2-(\bar\gamma\bar\nabla)^2]
\Phi^{\pm}_{\{n\}p_2i\alpha}=\varepsilon_{\{n\}}^2\Phi^{\pm}_{\{n\}p_
2i\alpha},
\]
one can easily conclude that in the basis
$\Phi^{\pm}_{\{n\}p_2i\alpha}$ the operator in the determinant
(\ref{eq.14}) is
diagonal.  Moreover,  its diagonal matrix elements are equal to
$4|\Delta|^2-p^2_0+(\mu\pm\varepsilon_{\{n\}})^2$.  Upon multiplying
these
quantities, one can find the determinant from (\ref{eq.14}).  In a
similar way, it is possible to calculate the first determinant from
(\ref{kp3_12}). Hence, taking into account the relation ${\rm
Det}O=\exp ({\rm Tr}\ln O)$, and following the standard procedure
(see, e.g., \cite{sch}), 
the following expression for $\tilde V$ is obtained from
(\ref{kp3_12})-(\ref{eq.14}) ( omitting an  infinite $\sigma$- and
$\Delta$-independent constant): 
\begin{eqnarray}\tilde
V&=&-\frac{\widetilde S}{v}=iN_f\int \frac{dp_0}{2\pi }
\biggl\{\sum_{\{p\}_0,\pm} \ln \left((E_{p}\pm\mu)^2-p_0^2\right)+
\nonumber\\
&+&A\sum_{\{n\},\pm}\ln
\left(4|\Delta|^2-p_0^2+(\varepsilon_{\{n\}}\pm\mu)^2\right)\biggr\},
\label{eq.15}
\end{eqnarray}
where $\{p\}_0$ denotes the set of quark quantum numbers for
vanishing background field
($\{p\}_0\equiv\{~-\infty<p_1,p_2,p_3<+\infty~\}$), and
$E_{p}=\sqrt{\bar p^2+\sigma^2}$. The factor $N_f$ in front of the
integral in (\ref{eq.15}) is the result of summation
over flavor indices $i=1,\dots,N_f$,
whereas the degeneracy factor $A\equiv gH/(8\pi^2)$ is due to
the integration over the momentum $p_2$ and summation over the color
indices $\alpha =1,2$.  Moreover, $\sum_{\{p\}_0}\equiv\int
d^3p/(2\pi)^3$, $\sum_{\{n\}}\equiv\int dp_3/(2\pi)\sum_{n,\zeta}$.

In the case of finite temperature $T= 1/\beta >0$
the corresponding expression for $\tilde V_T$  can be obtained
from (\ref{eq.15}) by means of the following replacements:
\[
\int\frac{dp_0}{2\pi }(\cdots)\to iT\sum_l(\cdots);~~~p_0\to
i\omega_l\equiv 2\pi iT(l+1/2);~~l=0,\pm 1,\pm 2,\ldots ,
\]
where $\omega_l$ is the Matsubara frequency. Hence,
\begin{eqnarray}
\tilde V_T&=&-N_fT\sum_{l=-\infty}^{l=\infty}
 \biggl\{\sum_{\{p\}_0,\pm}
 \ln \left((E_{p}\pm\mu)^2+\omega^2_l\right)+
 \nonumber\\
&+&A\sum_{\{n\},\pm}\ln
 \left(4|\Delta|^2+\omega^2_l+(\varepsilon_{\{n\}}\pm\mu)^2\right)
 \biggr\}.
\label{n15}
\end{eqnarray}

In order to transform (\ref{n15}), let us first perform the
summation over the Matsubara frequencies. It is evident that
\begin{equation}
  \sum_l\ln(\omega_l^2+\Omega^2)=
 \sum_l \int\limits_{1/\beta^2}^{\Omega^2}
 da^2\frac1{\omega_l^2+a^2}+
 \sum_l \ln\left(\omega_l^2+\frac1{\beta^2}\right),
\label{17}
\end{equation}
where $\Omega$ stands, according to (\ref{n15}), for
$\sqrt{(E_{p}\pm\mu)^2}$ or
$\sqrt{4|\Delta|^2+(\varepsilon_{\{n\}}\pm\mu)^2}$, i.e. $\Omega\geq
0$.
Note that we can  neglect the contribution from the last term in
(\ref{17}),
since it does not depend on $\sigma$ and $\Delta$.
 The first term in (\ref{17}) can be presented in the following
 form (see Appendix):
$$
  \sum_l\int\limits_{1/\beta^2}^{\Omega^2}
  da^2\frac1{\omega_l^2+a^2}=
  2\ln\ch(\Omega\beta/2)+\const
$$
\begin{equation}
  =\Omega\beta+2\ln\bigl(1+e^{-\Omega\beta}\bigr)+\const.
\label{20}
\end{equation}

Performing the summation over Matsubara frequencies in the
second term in 
(\ref{n15}), and taking into account the degeneracy of the quark
spectrum $\varepsilon_n$ in the chromomagnetic field 
with respect to
combination of quantum numbers $n$ and $\zeta$, we can use the
following expression for the energy spectrum:
$\varepsilon_n=\sqrt{gHn+p_3^2+\sigma^2}$, where
$n=0,1,2,\ldots$ is the Landau quantum number, and
$-\infty<p_3<\infty$. Then, summing over the spin quantum number
$\zeta=\pm1$, we have to account for the fact that for the ground
state with $n=0$ only one spin projection $\zeta =-1$ is possible.
Hence, a factor $\alpha_n=2-\delta_{n0}$ should be included in the
final expression. As for the summation over Matsubara frequencies in
the first term in (\ref{n15}), it is necessary to take into account
the
fact that the function (\ref{20}) is even with respect to the
variable
$\Omega$.  Finally, we thus arrive at the following result for the
thermodynamic potential 
\begin{eqnarray} && V_{H\mu
T}(\sigma,\Delta)=N_c\left(\frac{\sigma^2}{2G_{1}}+
\frac{|\Delta^2|}{G_2}
 \right)-\nonumber\\
 &-& 2N_f\int\frac{d^3p}{(2\pi)^3}(N_c-2)\biggl\{E_p+
 T\ln\Bigl[\bigl(1+e^{-\beta(E_p-\mu)}\bigr)
 \bigl(1+e^{-\beta(E_p+\mu)}\bigr)\Bigr]\biggr\}-\nonumber\\
   &-&N_fA\sum_{n=0}^\infty
  dp_3\alpha_n\biggl\{\sqrt{(\varepsilon_n-\mu)^2+4|\Delta|^2}
  +\sqrt{(\varepsilon_n+\mu)^2+4|\Delta|^2}+\nonumber\\
  &+&2T\ln\Bigl[\bigl(1+e^{-\beta\sqrt{(\varepsilon_n-\mu)^2+
  4|\Delta|^2}}\bigr)
 \bigl(1+e^{-\beta\sqrt{(\varepsilon_n+\mu)^2+4|\Delta|^2}}\bigr)
 \Bigr]\biggr\}.
 \label{eq.20}
\end{eqnarray}
For convenience, expressions are again written in terms of $N_f$
and $N_c$ even though in the following we will be concerned only with
$N_f=2$ and $N_c=3$.

\subsection{Regularization}

First of all, let us subtract from (\ref{eq.20}) an infinite
constant in order that the effective potential obeys the constraint
$V_{H\mu T}(0,0)=0$.
After this subtraction the effective potential still
remains
UV divergent. This divergency could evidently be removed by
introducing a simple momentum cutoff $|p|<\Lambda$. 
Instead of doing this, we find it convenient to use another
regularization procedure. To this end, let us recall that all UV
divergent contributions to the subtracted
potential $V_{H\mu T}(\sigma,\Delta)-V_{H\mu T}(0,0)$
are proportional to powers of meson and/or diquark fields
$\sigma$, $\Delta$.
So, one can insert some momentum-dependent form factors in
front of composite $\sigma$--and $\Delta$--fields in order to
regularize
the UV behaviour of integrals and sums.
\footnote
{A suitable physical motivation for such form factors follows
in the framework of nonlocal NJL type models based on the
one-gluon-exchange approximation
to $QCD$ with  nontrivial gluon propagator. In particular,
in ref.\cite{cahill} it was shown that the arising
exponential form factors for composite
mesons, as obtained from  the solution of the (nonlocal) 
Bete-Salpeter(BS)-equation,  make the
quark loop expansion including meson (diquark) vertices convergent. 
In this case, there is no need for introducing a sharp momentum
cutoff as in the local NJL model. Hence, the introduction of
smoothing form factors
in our expressions of an approximate local NJL model may be
interpreted as a
regularization procedure taking some effects of the 
originally non-local current-current interaction afterwards 
into account.} 

It is clear by now that we are going to study the effects of an
external chromomagnetic condensate field in the framework of the 
NJL-type model (1), which in addition to two independent 
coupling constants $G_1,G_2$ includes regularizing meson (diquark)
form factors. Of course, it
would be a very hard task
to study the competition of $\chi$SB and CSC for arbitrary values of
coupling constants $G_1,G_2$ and any form factors. 
Thus, in order to restrict this arbitrariness and to be able to
compare our results (at least roughly) with other approaches, 
we find it convenient to investigate  the phase structures
of the model (1) at $H=0$ and $H\ne 0$ only for some fixed values of
$G_1,G_2$ and some simple expressions for 
meson/diquark form factors (for simplicity, meson/diquark form
factors are chosen to be equal). 
Let us choose the form factors
\footnote{The application of the smooth meson
form factors (\ref{ff}) leads in a natural
way to a suppression of higher Landau levels, which is of particular
use here. Hence, this regularization scheme is particularly suitable
for the manifestation of the (chromo)magnetic catalysis 
effect of dynamical symmetry breaking.
Indeed, the (chromo)magnetic catalysis effect and
the underlying mechanism of dimensional reduction are closely related
to the infrared dominance of the lowest Landau level with $n=0$ \cite
{gus2,zheb}.} 
\begin{equation}
\phi=\frac{\Lambda^4}{(\Lambda^2+\vec p^{\kern2pt2})^2},\qquad
\phi_n=\frac
{\Lambda^4}{(\Lambda^2+p_3^2+gHn)^2},
\label{ff}
\end{equation}
which have to  be included in  the energy
spectra by a corresponding multiplication of the $\sigma-,\Delta-$
fields:
\begin{equation}
E_p^r=\sqrt {\vec p^{\kern2pt2}+\phi^2 \sigma^2}, \qquad
\varepsilon_n^r=\sqrt {gHn+p_3^2+\phi_n^2 \sigma^2},
\qquad |\Delta^2|\rightarrow \phi_n^2|\Delta^2|.
\label{22}
\end{equation}
Note that with the choice of simple form factors (\ref{ff})
our expression for the thermodynamic potential at
$H=0$ formally coincides with the corresponding expression of
Ref.\cite{berg} 
obtained for an NJL type model with instanton-induced four-fermion 
interactions. In particular, by a suitable choice of coupling
constants $G_1,G_2$,
we will later ``normalize'' our phase portraits for $H=0$ to the
curves
of this paper in order to illustrate the influence of a nonvanishing 
chromomagnetic field. 
\footnote{It is necessary to 
underline that in our case the meson/diquark form factors (\ref{ff})
mimick solutions of the BS-equation for some non-local
four-fermion interaction arising from the one-gluon exchange
approach to QCD. Contrary to this, the instanton-like form factor
used in \cite{berg} has another physical nature. It appears as  quark
zero
mode wave function in the presence of instantons \cite{rapp}.}

As a result, instead of (\ref{eq.20}) we shall deal with
the following regularized potential $V_r(\sigma,\Delta)$:
\begin{equation}
V_r (\sigma, \Delta)=V_0
-2N_\up {f} (N_\up {c} -2) \int\limits_{-\infty}^\infty V_1 \,\frac
{d^3p}{(2\pi)^3} -\frac {gHN_\up {f}} {8\pi^2}
\int\limits_{-\infty}^\infty \sum_{n=0}^\infty \alpha_n V_2 \, dp_3,
\label{21}
\end{equation}
where
$V_0=N_\up{c}\left( \frac {\sigma^2} {2G_1}+\frac {|\Delta^2|}
{G_2}\right)$,
\begin{eqnarray}
&&V_1=E^r_p+T\op {ln} \left [\left (1+e^{\ts -\beta (E^r_p+\mu)}
\right) \! \left (1+
e^{\ts -\beta (E^r_p-\mu)}\right )\right ],\nonumber\\
&&V_2=\sqrt {(\varepsilon_n^r -\mu)^2+4 |\Delta|^2\phi_n^2}
+ \sqrt {(\varepsilon_n^r+\mu)^2+4 |\Delta|^2\phi_n^2}+\\
&&+2T\op {ln} \left [\left (1+e^{\ts-\beta\sqrt
{(\varepsilon_n^r-\mu)^2+4|\Delta|^2\phi_n^2}} \right) \! \left
(1+e^{\ts-\beta\sqrt {(\varepsilon_n^r
+\mu)^2+4 |\Delta|^2\phi_n^2}} \right) \right]\nonumber
\end{eqnarray}
and $E_p^r$, $\varepsilon_n^r$ are given in (\ref{22}).
Despite the $\Lambda$-modification,
the expression (\ref{21}) contains yet UV-divergent integrals.
However, as it was pointed out from the very beginning, we
shall numerically study the subtracted effective
potential, i.e. the quantity $V_r(\sigma,\Delta)$--$V_r(0,0)$,
which has no divergences.

In the next section the dependency of the global minimum point
of the regularized potential (\ref{21}) on the external parameters
$H,\mu,T$ will be investigated.

\section{Numerical discussions}

In the previous section we have chosen the form factors as in
(\ref{ff}) in order to roughly normalize our numerical
calculations
at $H\ne 0$ on the results obtained at $H=0$ in \cite{berg}.
Comparing the effective potential (\ref{21}) at $gH=0$ with the
corresponding one from that paper
(denoting their respective diquark field and coupling constants by a
tilde), we see that 
these quantities coincide if $2\Delta=\tilde \Delta$,
$G_1=2N_c\tilde G_1$ and 
$G_2=N_c \tilde G_2$. 
Using further 
the numerical ratio of coupling constants from ref. \cite{berg}, 
we obtain  in our case the following relation: 
\begin{equation}
  G_2=\frac3{8} G_1.
\label{gg}
\end{equation}
Now, let us perform the numerical investigation of the global minimum
point (GMP) of the potential (\ref{21}) for form factors and values
of coupling constants as given by (\ref{ff}) and (\ref{gg}),
respectively. It was supposed earlier in some papers (see e.g.
\cite{sad}) that quantitative features of the color
superconducting phase transition 
might indeed depend on the value of the
form factor parameter $\Lambda$. 
So, in order to check the $\Lambda$-dependence of our results, 
we perform the
investigations for three different values of 
the cutoff,
$\Lambda=$0.6 GeV, 0.8 GeV and 1 GeV.
For each value of $\Lambda$, 
the corresponding value of $G_1$ is selected from the requirement
that the GMP of the function 
$V^r_{H\mu}(\sigma,\Delta)$ at $T=\mu=H=0$ is at the
point $\sigma = 0.4 $ GeV, $ \Delta=0$ in agreement with
phenomenological results and \cite{berg}. Then, the value of $G_2$
is fixed by the relation (\ref{gg}).
This yields, 
for example, $G_1\Lambda^2=2 N_\up{c}6.47$ at $\Lambda=0.8$ GeV,
$G_1\Lambda^2=2 N_\up{c}6.16$ at $\Lambda=1$ GeV etc.

Note, further that in the case of zero temperature we have
$$
V_1 |_{T=0}=E^r_p+(\mu-E^r_p) \, \theta (\mu-E^r_p),
$$
$$
V_2 |_{T=0}=\sqrt {(\varepsilon_n^r -\mu)^2+4 |\Delta|^2\phi_n^2}
+ \sqrt {(\varepsilon_n^r+\mu)^2+4|\Delta|^2\phi_n^2}.
$$
In order to study the phase structure of the model using numerical
methods, the summation over $n$ in (\ref{21}) is limited by a maximum
value $n_{max}=(2.5\Lambda)^2/gH$, where 
other terms of the series can be neglected due to their smallness.

First of all, it should be remarked that,
as in paper \cite{berg} at $gH=0$, a
mixed phase of the model
was not found for $H\ne 0$, i.e. for a wide range of parameters
$\mu,H,T$ we
did not find a global minimum point of the potential
(\ref{21}), at which $\sigma\ne 0$, $\Delta\ne 0$. The results of
our numerical investigations of the GMP of $V_r (\sigma,\Delta)$ are
graphically represented  in the set of 
Figs. 1-6, where the notations
I, II and III are used for the symmetric phase, for the phase with 
chiral symmetry breaking and for the CSC phase, respectively.
For the points from region I the GMP of the potential lies at
$\sigma=0$, $\Delta=0$. In  region II we have a phase with broken
chiral symmetry,  corresponding to the GMP of the
potential at $\sigma\ne 0$, $\Delta=0$. Finally, the color
superconducting
phase with the GMP of the potential at $\sigma=0$, $\Delta\ne 0$,
corresponds to the points from region III of these figures.

In  Fig. 1, one can see the phase portrait of the model
in terms of $\mu, gH$ at $T=0$ for each of the above mentioned values
of the cutoff $\Lambda$.
 The boundary between phases III and II is practically 
$\Lambda$-independent and represents a first order phase
transition curve $\mu_{cr}(gH)$. It is necessary to note also that
for each value of $\Lambda$ and 
a fixed value of $gH$ there is a critical chemical potential
$\tilde\mu_c(H)$, at which the GMP is transformed from a point of
type III
to a symmetric point of type I. However, this phase transition is 
remarkably $\Lambda$-dependent. Indeed, even in the simplest
case with $H=0$ we have 
 $\tilde\mu_c(0)=1$ GeV at $\Lambda=0.6$ GeV,
$\tilde\mu_c(0)=1.3$ GeV at $\Lambda=0.8$ GeV,
$\tilde\mu_c(0)=1.65$ GeV at $\Lambda=1$ GeV. 
It is well-known from 
the one-gluon exchange approximation in QCD
\cite{son,raj} that
CSC can exist even at enormously  high
values of the
chemical potential $\mu\gtrsim 10^8$ MeV. So the
above mentioned 
NJL framed transition from 
phase III to phase I looks, in the case $T=0$, 
rather like
an artefact of the regularization procedure.
\footnote{Note that in \cite{sad}
it was also claimed that in 
the NJL model at 
high enough $\mu$ the diquark condensate vanishes, which is the 
consequence of the regularization by 
a form factor.  In this region, it might be 
necessary to use another approximation for the CSC investigation.}
Thus, since this phase transition turns out to be 
unphysical, it is not shown in Fig. 1.
The critical curves of this figure are obtained 
by interpolation in the most simple manner, i.e. by a
second order polynomial, of numerical points lying at $gH>0.2$
GeV$^2$ 
(for technical reasons, such points are explicitely 
shown in Fig.2, rather than in Fig.1)
Earlier, in the papers \cite{ekvv} the model (1) at $G_2=0$
and in the presence of an external magnetic field was considered.
As shown there, for small values of the magnetic field strength
the critical curves, as well as various thermodynamical
and dynamical parameters of the system, demonstrate oscillating
behaviour.
In order to make more accurate interpolations and to
become sure whether analogous oscillations appear in the present
case  (i.e., for the critical curves in
Figs.1, 2) as well, or do not, one should make an enormous 
amount of numerical
calculations, which proved to be rather difficult to accomplish. Due
to this, we can make only a
conjecture of an oscillating behavior of the critical curves
judging from the positions of the points we have really calculated.
Note further that in the region of low chromomagnetic fields,
$gH<0.2$ GeV$^2$, we have extrapolated the critical curves to the
known points at $gH=0$.

In  Fig. 2, the phase portrait of the model
in terms of $\mu, gH$ at $T=0.15$ GeV is presented 
for each value of parameter $\Lambda=$0.6 GeV, 0.8 GeV, and 1 GeV.
One can see that the boundary between II and III phases 
(a critical curve of a first order phase transition)
 only slowly changes with varying $\Lambda$. 
The second order phase transition from 
the CSC phase III to the symmetric phase I,
which for finite T is now supposed to really exist,
also has
a weak $\Lambda$-dependence for $gH<0.2$ GeV$^2$.
At greater values of $gH$ the boundary 
between III and I phases has, however,
a stronger $\Lambda$-dependence. 
The phase diagrams in  the $(T, \mu)$ plane  for $gH=0$ and $gH=0.4$
GeV$^2$ are schematically represented in Fig. 3. The phase diagram in
the $(T,gH)$ plane  for $\mu=0.4$ GeV  is represented in Fig. 4. 
For both figures we choose $\Lambda=0.8$ GeV,  for simplicity. It
is necessary to point out that at $gH=0$ the numerical results of
Figs. 1--4 coincide with those obtained in \cite{berg} at
$\Lambda=0.8$ GeV. Moreover, it should be emphasized that in all the
above mentioned figures, a second order phase transition takes
place at the boundary of the region I. At the boundary between
regions II and III a first order phase transition takes place.

Let us for a moment fix the value of the chemical potential and
temperature at varying values of 
$gH$. In this case, in 
the Figs. 1, 2, one will have a straight line 
parallel to
the $gH$ axis. In particular, if $\mu=0.4$ GeV and $T=0$, 
in Fig.1 this line originates 
at $gH=0$ in the CSC phase III. At some (critical) value
$(gH)_c\approx
0.1$ GeV$^2$ it crosses the line $\mu_{cr}(gH)$ and then, at yet
greater values of $gH$, it passes through the phase II.
Accordingly, at $gH<(gH)_c$ the GMP of the effective potential lies
at the point $(\sigma=0,\Delta\ne 0)$, where $\Delta$ is equal to the 
diquark condensate in the true stable vacuum, 
whereas at $gH>(gH)_c$ the
point $(0,\Delta\ne 0)$ ceases to be a GMP. In this case it is only a
local minimum point, so that $\Delta\ne 0$ 
corresponds to a metastable ground state of the
system (for the stable ground state at $gH>(gH)_c$ 
the GMP is of the form $\sigma\ne 0,\Delta =0$).
Thus, the value $(gH)_c$ is the so-called evaporation point for the
diquark condensate. 

In Fig. 5, the diquark condensate $\Delta$ is depicted as a function
of
$gH$ for three values of the temperature at $\mu=0.4$ GeV
and $\Lambda=0.8$ GeV (in this Figure, due to problems with
distinguishing 
closely positioned points for different cutoff values, we restricted 
ourselves to plotting curves for only one cutoff value $\Lambda =0.8$
GeV.).
Thick curves 
correspond to a stable diquark condensate (the point 
$(0,\Delta\ne 0)$ is the global minimum
 of the effective potential), and thin curves correspond to a
 quasistable diquark condensate (this  is a local minimum). 

Here we should note that
recent investigations yield the following value of 
the QCD gluon condensate at $T=\mu=0$:
$gH\approx 0.6$ GeV$^2$ \cite{glue}. 
It was shown in \cite{saito} that
in the framework of
a quark-meson model  at ordinary nuclear density
$\rho_0$ the gluon condensate decreases by no more than six percent,
compared with its value at zero density. At densities $3\rho_0$ the
value of  $\langle FF\rangle$ decreases by fifteen percent. This
means that for values
of the chemical potential $\mu <1$ GeV the gluon condensate is 
a slowly decreasing function vs $\mu$. Taking in mind this 
fact, one can draw  an important  conclusion from our
numerical analysis: At $H=0$, $T=0$  and $\mu=0.4$ GeV there 
should exist the CSC phase (see \cite{berg}). However, if the 
condensate value
$gH\approx 0.5$ GeV$^2$ is taken into account at $\mu=0.4$ GeV,
then our model consideration concludes that the CSC phase does not
exist at $T=0$ for such a large value of the gluon condensate
(cf. Figs. 4, 5).
\footnote{Recently, a similar prediction, namely that
nonperturbative gluon fluctuations might be strong enough to 
destroy the CSC, was done in \cite{agas}, but in a rather 
qualitative form.}
Assuming that our results would remain valid also for more
realistic condensate fields, this would seemingly render it
difficult to observe  a CSC phase for $T=0$
at $\mu=0.4$ GeV.

Notice, however, that our results change at finite temperature.
First, if one takes $T=0.1$ GeV and $\mu=0.4$ GeV, then at
sufficiently 
small values of $gH$ there is a symmetric phase of the theory, where
both $<\bar qq>$ and $<qq>$ condensates are zero (see Fig. 5).
However, at the point $gH\approx 0.1$ GeV$^2$ there is a second order
phase
transition to the CSC phase (here only $<qq>\ne 0$), and at the 
point $gH\approx 0.2$ GeV$^2$ the external field destroys the 
CSC in favour of the chiral phase, where $<\bar qq>\ne 0$ and
$<qq>=0$.
A similar behaviour is observed at $T=0.15$ GeV and $\mu=0.4$ GeV
(see Fig. 5).
Secondly, if one takes $T=0.15$ GeV and $\mu\geq 0.6$ GeV, then at
sufficiently 
small values of $gH$ there is a symmetric phase of the theory, where
both $<\bar qq>$ and $<qq>$ condensates are zero (see Fig. 2).
However, at the point $gH\geq 0.3$ GeV$^2$ there is a second order
phase
transition to the CSC phase, where only $<qq>\ne 0$.
Hence, in some cases the gluon condensate can induce the CSC
(the so-called chromomagnetic catalysis effect of CSC).

This effect is further
illustrated in  Fig. 6, where the diquark
condensate vs $gH$ is depicted at $\mu=0.8$ GeV for two
temperatures $T=0$, $T=0.15$ GeV and
two values of $\Lambda=0.8$ GeV and $\Lambda=1$ GeV.
One can easily see that the value of $\Delta(gH)$ for
$T=0.15$ GeV is identically zero for
$gH\leq 0.44$ GeV$^2$ with $\Lambda =0.8$ Gev, and  for
$gH\leq 0.52$ GeV$^2$ with $\Lambda =1$ Gev, i.e.,
for each value of $\Lambda$,  there exists a critical value of 
$gH$ at which the nonzero diquark condensate is generated.

Finally, let us make some additional remarks concerning the diquark 
condensate at $\mu=0.8$ GeV. As it follows
from our numerical analysis at $T=0$ (see Fig. 1, 6),
for $\mu=0.8$ GeV, $gH=0$,  the GMP of the effective potential
corresponds to the CSC phase with a stable diquark condensate
$\Delta\leq 0.1$ GeV. However, assuming that 
the value of the gluon condensate $gH\approx 0.4$
GeV$^2$ would hold for the above nonvanishing chemical
potential,
one would get a value of the  diquark condensate $\Delta\gtrsim 0.2$
GeV, which is significantly larger in magnitude, than at
$gH=0$. It follows from Fig. 6 that the diquark condensate noticeably
depends on $\Lambda$. One could suppose that 
this is due to the rather high value of the considered
chemical potential, $\mu=0.8$ GeV. However,
as our calculations show, a similar 
$\Lambda$-dependence of the diquark condensate 
occurs for  smaller values of $\mu$ as well. These 
results confirm the conclusions of the paper \cite{sad} 
obtained at $H=0$ that the value of the diquark
condensate varies with $\Lambda$, if the form factor regularization
is used in the NJL model. In contrast, the points of the boundary
between II and III phases do not show  a remarkable
$\Lambda$-dependence (see Fig. 1,2).

As a general conclusion, we see that taking into account an
external chromomagnetic field
at least in the form as considered in the model above,
might, in principle, lead to  remarkable
qualitative and quantitative changes in the 
picture of the
diquark condensate formation, obtained in the framework of 
NJL models at $H=0$. 
This concerns, in particular, the possibility of a transition to 
the CSC phase and diquark condensation at finite temperature.
Clearly, a detailed quantitative
discussion would, however, require to have 
additional information on the
gluon condensate as a function of the temperature and chemical
potential and on a possible $\mu$-and $gH$-dependence of
the cutoff parameter $\Lambda$. 
\footnote{Such a dependence might, for example, arise, if one
identifies the cutoff in the local 4-fermion model with 
the effective gluon mass in the gluon propagator.}
A further interesting generalization could be
to extend this kind of approach to inhomogeneous background field
configurations \cite {cal}. This concerns, in particular, the
nonabelian condensate fields of the
Stochastic Vacuum Model of QCD realizing Wilson's area law of 
confinement \cite {sim}.
 
\section{Summary and conclusions}

In the present paper the influence of different physical factors on
the phase structure of the two-flavor NJL model (1) with two
independent structures of four-quark interactions has been
considered. This model is
adequate for the description of the low energy physics of 
two-flavor QCD both in $q\bar q$- and $qq$ channels. In the papers
\cite{alf}-\cite{klev} it was shown  that in
QCD-motivated type of models (1) with $H=0$ the new color
superconducting (CSC) phase can exist for  
moderate values of the chemical potential (baryon density). As
generalization
of the ``free field'' NJL model of ref. 
\cite{klev}, we have, in particular, taken into account such
nonperturbative
feature of the real QCD vacuum as the nonzero gluon chromomagnetic
condensate $<F^a_{\mu\nu}F^{a\mu\nu}>$$\equiv 2H^2$, which in the
framework of a NJL model can be simulated 
by an external chromomagnetic field. The recent
estimates  give the following fixed
value $gH_{phys}\approx  0.6$ GeV$^2$ \cite{glue} for the gluon
condensate in QCD at $\mu,T=0$. Despite of this fact, we considered
it, however, useful to treat $H$ as a free external parameter of the
model.

Since in the CSC phase the original $SU_c$(3) symmetry of the theory 
is spontaneously broken down to $SU_c$(2), five color gauge
bosons acquire masses. The corresponding external fields are expelled
from the CSC phase (Meissner effect). However, the other three
``color isospin'' gauge bosons remain massless,  in accordance with
the
residual $SU_c$(2)
symmetry of the vacuum. Clearly,
the corresponding external fields may then penetrate into the CSC
phase.
It is just the influence of these types of external chromomagnetic
fields 
on the formation of CSC which was studied in our previous paper
\cite{ebklim} in the framework of a (2+1)-dimensional NJL model 
for vanishing chemical potential and temperature.
There it was shown that chromomagnetic fields may induce the CSC
phase
transition. In the present paper, the chromomagnetic generation of
CSC has been studied  in the 
framework of a (3+1)-dimensional NJL model for finite 
chemical potential
and temperature. The vector-potential of the external chromomagnetic
field was chosen to be of the Matinyan-Savvidy form (\ref{x7}) and
lies in the algebra of the residual $SU_c$(2) group, too.

The coupling constants $G_1,G_2$ of our model (1) are considered as
free independent
parameters. In our numerical estimates, we found it 
convenient to use the relation (\ref{gg})
in order to ``normalize'' our calculations at $H\ne 0$ on the known
results at $H=0$ \cite{berg}. However, we hope that our
qualitative conclusions remain also valid for values of $G_1,G_2$ in
some neighbourhood of (\ref{gg}).
The results of numerical investigations of the effective potential 
(\ref{21}) are presented in a set of Figs 1--6, where phase
diagrams for the extended NJL model (1) in terms of $\mu,T,H$ as well
as the behaviour of diquark condensates versus $gH$ are shown.

First of all we should note that the form factor regularization of
the NJL model (1) is used throughout in the present paper.
As shown in \cite{sad}, in this case even at $H=0$ the features
of CSC depend on the value of the form factor
parameter $\Lambda$.  So, in order to clarify the 
corresponding situation at $H\ne
0$,  we have used
three different values of $\Lambda=$0.6 GeV, 0.8 GeV, and 1 GeV 
in the regularization scheme under consideration. 
One can see that 
phase transitions between 
the chirally broken and CSC phases have 
rather weak $\Lambda$-dependence (see Figs 1,2), whereas the 
diquark condensate values noticeably depend on $\Lambda$
(see, e.g. Fig. 6).

The main conclusion of our investigations is that the inclusion of an
external chromomagnetic field can significantly change the
phase portrait, obtained at $H=0$.
Indeed, at $H=T=0$ the values of the 
chemical potential corresponding to the CSC phase approximately lie
in the interval $\mu >0.3$ GeV 
(see Fig. 1).
If the external chromomagnetic field of the type  (\ref{x7})
is switched on at $\mu=0.4$ GeV, then at $gH_c\approx 0.1$ GeV$^2$ 
there is a transition of the system
from CSC to a phase, where only chiral symmetry is broken
down. Thus, at $gH_{phys}\approx  0.6$ GeV$^2$, $\mu=0.4$ GeV and
$T=0$
the CSC can not be observed at all. However, if $T=0$ and the
chemical
potential is fixed at $\mu=0.8$ GeV, then at least for all values 
$0\leq gH\leq 0.6$ GeV$^2$ one can observe the CSC phase in which
the diquark condensate $\Delta(H)$ is nonzero. It is worth
mentioning that in this case the function $\Delta(H)$ is 
monotonically increasing (see Fig. 6) and the value of the
diquark
condensate at $gH_{phys}$ turns out to be significantly greater 
in magnitude than at vanishing $H$.
\footnote{We roughly suppose throughout the present paper that
at $(\mu,T)\ne 0$ the real gluon condensate is the same as at
$(\mu,T)=0$. 
However, using a given $\mu,T$-dependency
of the gluon condensate, it would be possible to extract physical
informations about the CSC phase using 
our phase diagrams in Figs. 1-4 and plots of $\Delta(H)$
functions in Figs. 5, 6.}

Finally, one should note that at $\mu\ne 0, T\ne 0$
the external
chromomagnetic field can induce the CSC phase transition.
For example, at $T=0.15$ GeV and $\mu=0.8$ GeV there is a symmetric
phase of the theory in which $\sigma=\Delta=0$ (both chiral and
diquark condensates are zero) for sufficiently small values of $gH$
(see Fig. 2).
However, at some critical point $gH_c$,
a phase transition of the second order from the symmetric to the 
CSC phase is induced by the external chromomagnetic field.
Notice that the CSC induction by some types of external
chromomagnetic
fields was observed in the framework of a NJL model  at zero $\mu$,
$T$ (see \cite{ebklim}-\cite{gap1}). The
present analysis shows that this effect takes place at some nonzero
values of $\mu,T$, too, 
which in principle could be important for
heavy ion-ion collisions taking place at nonzero temperature.
Clearly, a somewhat unpleasant feature of the above NJL approach, 
which requires some caution, is the 
$\Lambda$ dependence of some of the results. 
Nevertheless, we believe that the above results are interesting 
and may serve as a starting point for further 
investigations of this issue.

\section*{Acknowledgements}
We wish to thank V.P. Gusynin, V.A. Miransky, Y. Nambu and H. Toki
for fruitful discussions. D.E. gratefully acknowledges the support
provided to him by the Ministry of Education and Science and
Technology of Japan (Monkasho) for his work at RCNP
of Osaka University.
This work is supported in part  by DFG-Project 436 RUS 113/477/4.

\appendix
\section*{}
Let us sketch the calculation of the integral in eq. (\ref{20}) of
the text. Evidently, it can be rewritten as a contour integral
\begin{equation}
  \sum_l\int\limits_{1/\beta^2}^{\Omega^2}
  da^2\frac1{\omega_l^2+a^2}=
  -\int\limits_{1/\beta^2}^{\Omega^2} da^2
  \int\limits_{C_0}\frac{d\omega}{2\pi
  i}\frac1{\omega^2+a^2}\frac\beta
  2\tg  \frac{\beta \omega}{2},
  \label{A1}
\end{equation}
where $\tg(\beta\omega/2)$ has poles inside the integration
path
$C_0$ (see Fig. 7):
\begin{equation}
  \frac{\beta\omega}{2}=\pm\frac\pi2(2l+1),\quad l=0, 1,2, \ldots
\end{equation}

The integral over $C_0$ is equal to the integral along the contour
$C$ (Fig. 8), and hence
$$ -\int\limits_{C_0}\frac{d\omega}{2\pi
  i}\int\limits^{\Omega^2}_{1/\beta^2}
  \frac{da^2}{a^2+\omega^2}\frac\beta2 \tg\frac{\beta\omega}2
$$
$$
  =-\int\limits^{\Omega^2}_{1/\beta^2}da^2\int\limits_C
  \frac{d\omega}{2\pi i }\frac{\beta}{2}
  \left(\frac1{\omega-ia}-\frac1{\omega+ia}\right)\frac1{2ia}\tg\frac
  {\beta\omega}{2}
$$
\begin{equation}
  =\beta\int\limits^{\Omega}_{1/\beta} da\th\frac{\beta
  a}2=2\ln\ch\frac{\beta a}2
  \biggr|^{a=\Omega}_{a=1/\beta}
  =2\ln\ch\frac{\Omega\beta}2+\const.
\end{equation}

\newpage

\begin{center}
{\bf Figure captions}
\end{center}
\vspace*{0.5cm}

{\bf Fig. 1.}~~ Phase portrait of the model in terms of variables
$(\mu,gH)$ at $T=0$ for three values of the cutoff parameter
$\Lambda=
0.6, 0.8, 1$ GeV. Regions II, III describe the 
phase with broken chiral symmetry 
($\sigma\ne 0, \Delta=0$) and the color superconducting phase 
($\sigma=0,\Delta\ne 0$), respectively.    
\vspace{0.5cm}

{\bf Fig. 2.}~~Phase portrait of the model in terms of variables
$(\mu,gH)$ at $T=0.15$ GeV for three cutoff values $\Lambda=
0.6, 0.8, 1$ GeV.
The included phase I is the symmetrical one ($\sigma=0,\Delta=0$).
\vspace{0.5cm}

{\bf Fig. 3.}~~Phase portrait of the model in terms of variables
$(T,\mu)$ at $gH=0$ and $gH=0.4$ GeV$^2$ for $\Lambda=
0.8$ GeV. 
\vspace{0.5cm}

{\bf Fig. 4.}~~Phase portrait of the model in terms of variables
$(gH,T)$ at $\mu=0.4$ GeV and $\Lambda=0.8$ GeV .
\vspace{0.5cm}

{\bf Fig. 5.}~~Diquark condensate as a function of an external
chromomagnetic field for $\mu=0.4$ GeV, $\Lambda=0.8$ GeV and 
three values of the temperature, $T=0$, $T=0.1$ 
and $T=0.15$ GeV.
\vspace{0.5cm}

{\bf Fig. 6.}~~Diquark condensate as a function of an external
chromomagnetic field for 
$\mu=0.8$ GeV 
and two values of the temperature, $T=0$, 
and $T=0.15$ GeV and for two values of the cutoff parameter
$\Lambda= 0.8, 1$ GeV.
\vspace{0.5cm}

{\bf Fig. 7.}~~Integration path $C_0$ used in eq. (A1) of Appendix.
\vspace{0.5cm}

{\bf Fig. 8.}~~Integration path $C$ used in eq. (A3) of Appendix.

\newpage
%%%%%%%%%%%%%%%%%%%%%Figures%%%%%%%%%%%%%%%%%
%\vspace{2cm}

\begin{figure}\centering
\epsfxsize=80mm\epsfbox{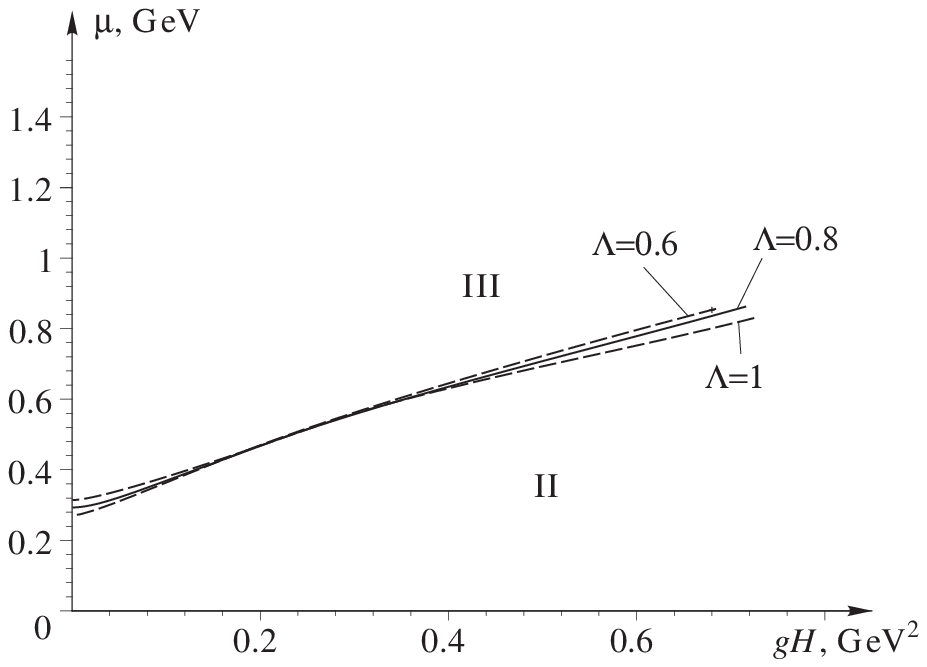}
\caption{}
\end{figure}

\begin{figure}\centering
\epsfxsize=80mm\epsfbox{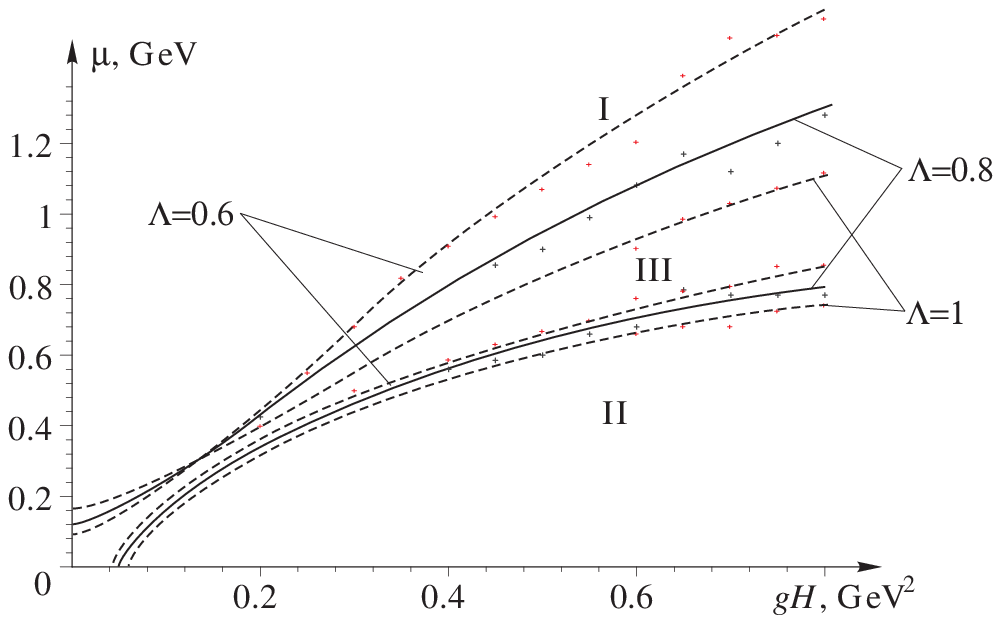}
\caption{}
\end{figure}

\begin{figure}\centering\epsfxsize=80mm
\mbox{\epsfbox{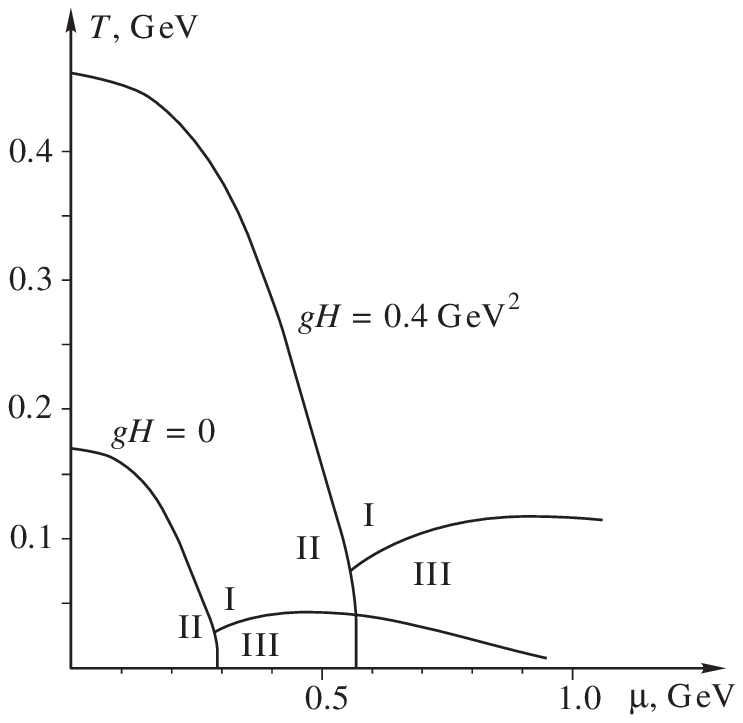}}
\caption{}
\end{figure}

\begin{figure}\centering
\mbox{\epsfbox{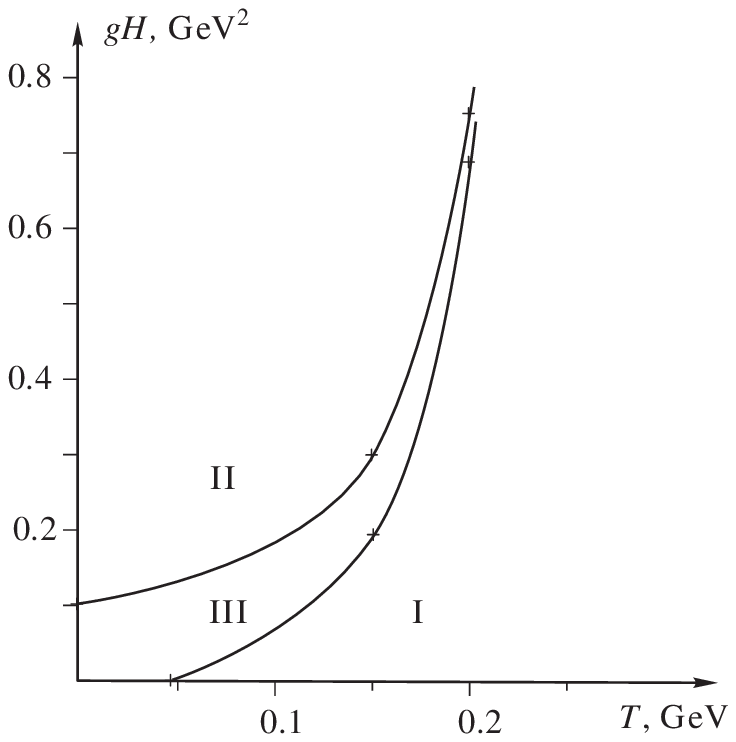}}
\caption{}
\end{figure}

\begin{figure}\centering
\mbox{\epsfbox{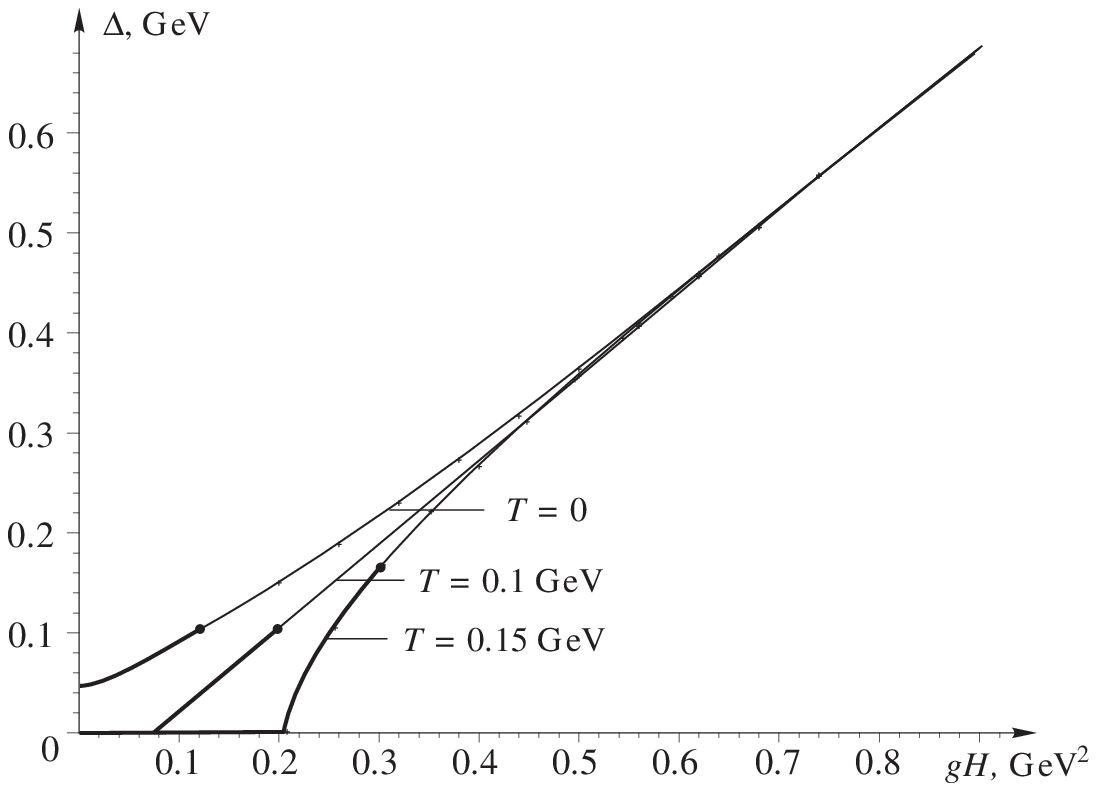}}
\caption{}
\end{figure}

\begin{figure}\centering
\mbox{\epsfbox{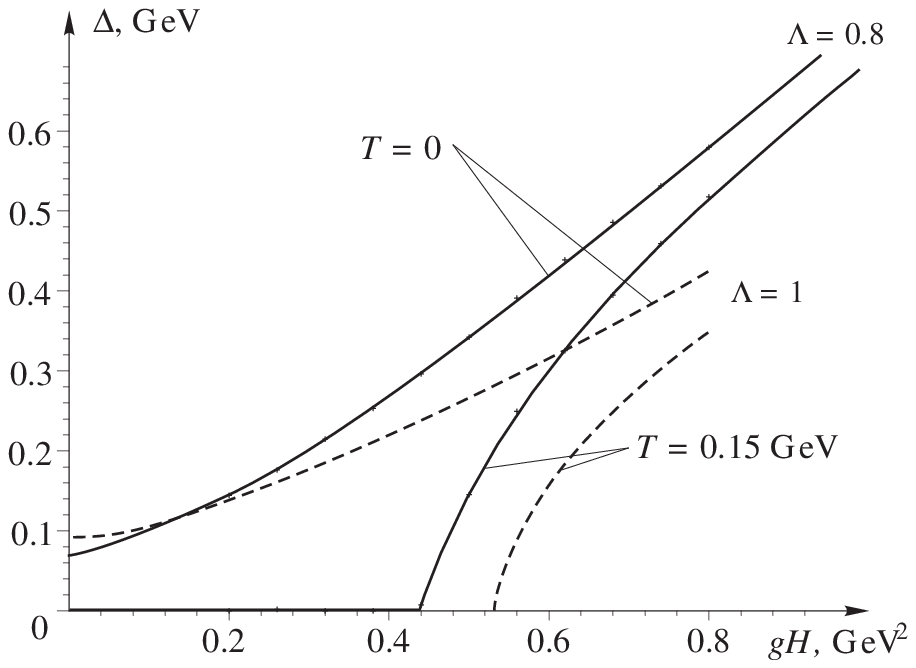}}
\caption{}
\end{figure}

\begin{figure}\centering
\mbox{\epsfbox{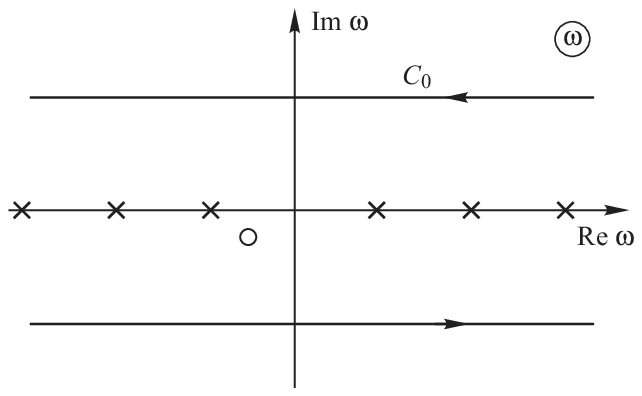}}
\caption{}
\end{figure}

\begin{figure}\centering
\mbox{\epsfbox{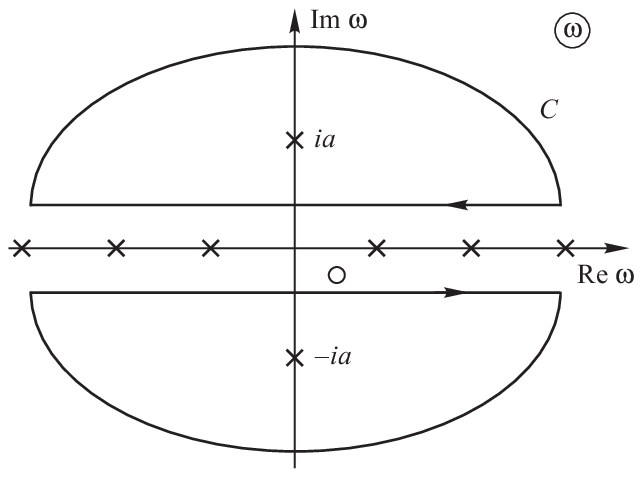}}
\caption{}
\end{figure}

\end{document}